\input{aipcheck}

\documentclass[final]{aipproc}
\layoutstyle{6x9}

\def\ga{\mathrel{\raise.3ex\hbox{$>$\kern-.75em\lower1ex\hbox{$\sim$}}}}
\def\la{\mathrel{\raise.3ex\hbox{$<$\kern-.75em\lower1ex\hbox{$\sim$}}}}
\def\beq{\begin{equation}}
\def\eeq{\end{equation}}
\newcommand{\gev}{\,\, \mathrm{GeV}}
\newcommand{\tb}{\tan\beta}
\newcommand{\Mh}{M_h}
\newcommand{\ssi}{\sigma^{\rm SI}_p}
\newcommand{\neu}[1]{\tilde \chi^0_{#1}}
\newcommand{\mneu}[1]{m_{\tilde \chi^0_{#1}}}
\def\calh{\mathcal{H}}
\def\nl{\hfill\nonumber\\&&}
\def\nnl{\hfill\nonumber\\}
\def\mfiv{m_\mathbf{\overline{5}}}
\def\mfivl{m_{\mathbf{\overline{5}},1}}
\def\mten{m_\mathbf{10}}
\def\mtenl{m_{\mathbf{10},1}}
\def\afiv{A_\mathbf{\overline{5}}}
\def\aten{A_\mathbf{10}}
\def\ohsq{\Omega_{\chi} h^2}

\begin{document}

\title{Dark Matter in SuperGUT Unification Models$^*$}

\classification{}
\keywords{}

\author{Keith A. Olive}{
  address={William I. Fine Theoretical Physics Institute, \\
  University of Minnesota, Minneapolis, MN 55455, USA \\
  $^*$To be published in the Proceedings of the 6th DSU Conference, Leon, Mexico, ed. D. Delepine}
}

%\author{<author2>}{
%  address={<common address for author2 and author3>}
%}

%\author{<author3>}{
%  address={<common address for author2 and author3>}
%  ,altaddress={<author1 address>} % additional visiting address
%}

\vskip - 2.5in
\rightline{UMN--TH--2913/10}
\rightline{FTPI--MINN--10/21}
%\rightline{August 2010}
\vskip -.4in
\begin{abstract}
After a brief update on the prospects for dark matter
in the constrained version of the MSSM (CMSSM) and its differences
with models based on minimal supergravity (mSUGRA),
I will consider the effects of unifying the supersymmetry-breaking
parameters at a scale above $M_{GUT}$. One of the consequences
of superGUT unification, is the ability to 
take vanishing scalar masses at the unification scale with
a neutralino LSP dark matter candidate.
This allows one to resurrect no-scale supergravity as a viable 
phenomenological model.

\end{abstract}

\maketitle

%%%%%%%%%%%%%%%%%%%%%%%%%%%%%%%%%%%%%%%%%%%%
%% MAINMATTER
%%%%%%%%%%%%%%%%%%%%%%%%%%%%%%%%%%%%%%%%%%%%

\section{Introduction}

 While often used synonymously, the constrained minimal supersymmetric
 standard model (CMSSM) \cite{funnel,cmssm,efgosi,cmssmwmap,like1,like2} 
 differs in two important ways from supersymmetric models
 based on minimal supergravity (mSUGRA) \cite{Polonyi,Fetal,BIM,bfs}. The latter class of theories
 is in fact a subset of the former and can be thought of as a very constrained version of the theory \cite{vcmssm}.
 The often studied CMSSM, is a 4+ parameter theory. Starting with the superpotential,
 \beq
W =  \bigl[ y_e H_1  L e^c + y_d H_1 Q d^c + y_u H_2
Q u^c \bigr] +\mu H_1 H_2 ,
\label{suppot}
\eeq 
we can obtain the soft supersymmetry-breaking part of the Lagrangian
\begin{eqnarray}
\mathcal{L}_{\rm soft} & =  & - \frac{1}{2} M_\alpha \lambda^\alpha \lambda^\alpha - 
m^2_{ij} {\phi^*}^i \phi^j \\ \nonumber
& & - A_e y_e H_1  L e^c - A_d y_d H_1 Q d^c - A_u y_u H_2
Q u^c - B \mu H_1 H_2 .
\end{eqnarray}
In the CMSSM, the four parameters are:  the gaugino mass, $M_\alpha = m_{1/2}$, unified
at some high energy input scale, $M_{in}$, 
usually assumed to be the grand unified (GUT) scale, $M_{GUT}$;
the scalar masses $m^2_{ij}= \delta_{ij}m^2_0$; and the trilinear terms, $A_f = A_0$, 
are all unified at the same input 
scale, $M_{in}$; and the ratio of the two Higgs expectation values, $\tan \beta$. 
In the CMSSM, $|\mu|$ and $B$ are determined by the electroweak symmetry-breaking (EWSB)
conditions by fixing $M_Z$ and $\tan \beta$. 
In mSUGRA models, there is an additional constraint, namely 
\beq 
B_0 = A_0 - m_0 .
\label{msug}
\eeq
In this case, it is no longer possible to satisfy the EWSB conditions
(i.e., the minimization of the Higgs potential) and choose $\tan \beta$ independently.  As a consequence, we have a 3+ parameter theory
which is now specified by $m_{1/2}, m_0$, and $A_0$.
The "+" in the number of parameters refers to the sign of $\mu$. While
the magnitude of $\mu$ is fixed by the EWSB conditions, its sign is
left undetermined.
 In true minimal supergravity models,
there is also a relation between $m_0$ and the gravitino mass, namely $m_0 = m_{3/2}$.
Thus, the gravitino mass is no longer independent and as we will see, often in 
mSUGRA models, the gravitino is found to be the lightest supersymmetric particle (LSP).
The gravitino mass is generally not counted as one of the parameters in the CMSSM, 
as it can be set
to a sufficiently large value so as to render it irrelevant.

No-scale supergravity models \cite{nosc1} are still more restrictive, as the boundary conditions
on the scalar masses and bi- and trilinear supersymmetry-breaking terms become
\beq
m_0 = A_0 = B_0 = 0 .
\label{nosc}
\eeq
Thus, we are left with a 1+ parameter theory specified by $m_{1/2}$. 
Generally, the condition $m_0 = 0$ is largely incompatible with phenomenology 
except for a restricted set of values of $\tan \beta$, but might be more viable
 if the supersymmetry-breaking scale were pushed above the GUT scale \cite{eno}.
 
 Below, I will try to highlight some of the differences between the CMSSM and mSUGRA models
 and explore the phenomenological consequences of pushing the scale at which the
 supersymmetry-breaking parameters are unified above the GUT scale \cite{emo,emo2}.
 After a brief review on the status of the CMSSM, and its differences with mSUGRA,
 I will demonstrate the difficulties with no-scale supergravity models.
 Then, I will describe the effects of raising the supersymmetry-breaking scale above
 the GUT scale and  apply this to no-scale models.

\section{The CMSSM}

\begin{figure}
  \includegraphics[width=.45\textwidth]{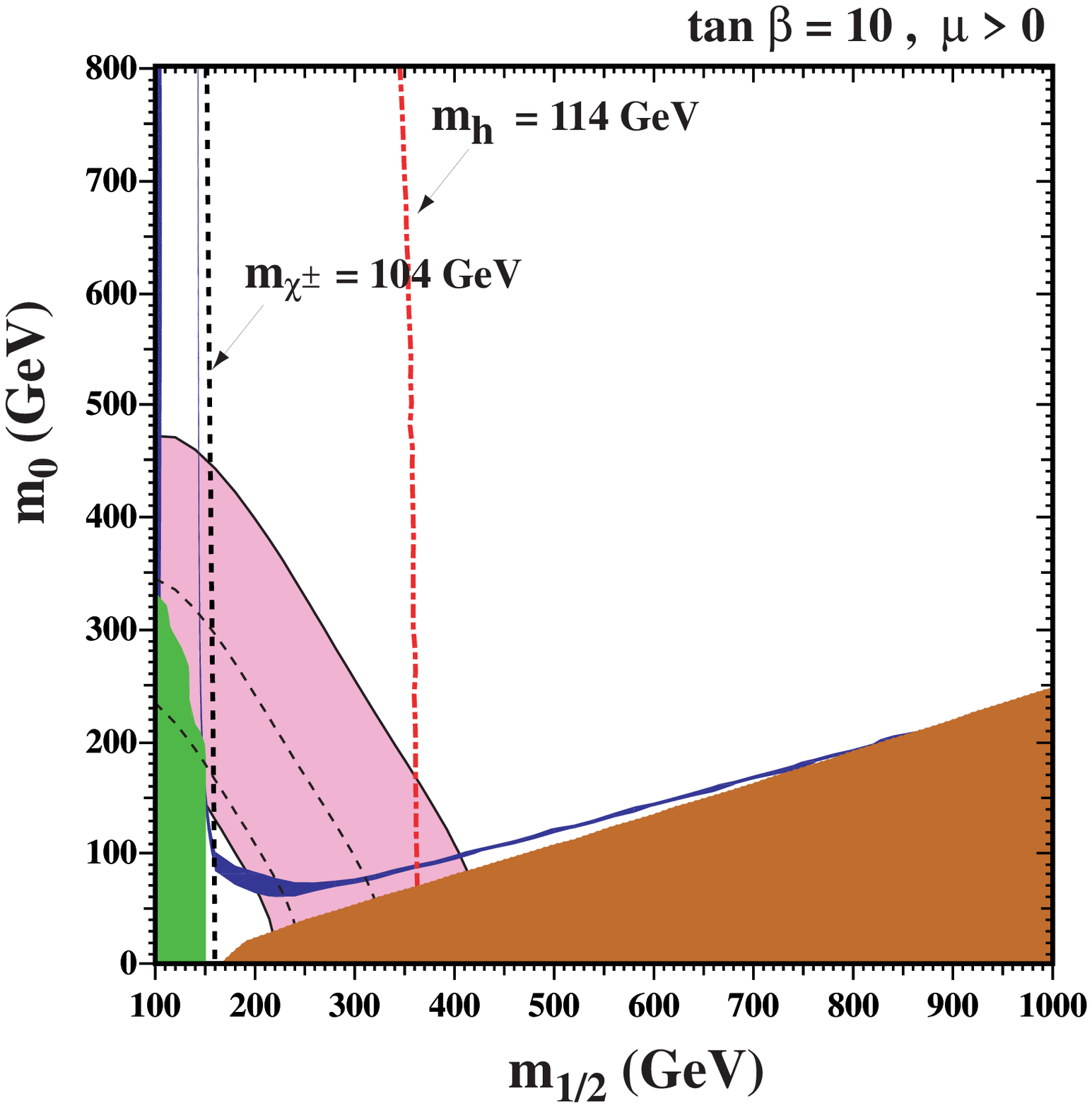}
    \includegraphics[width=.45\textwidth]{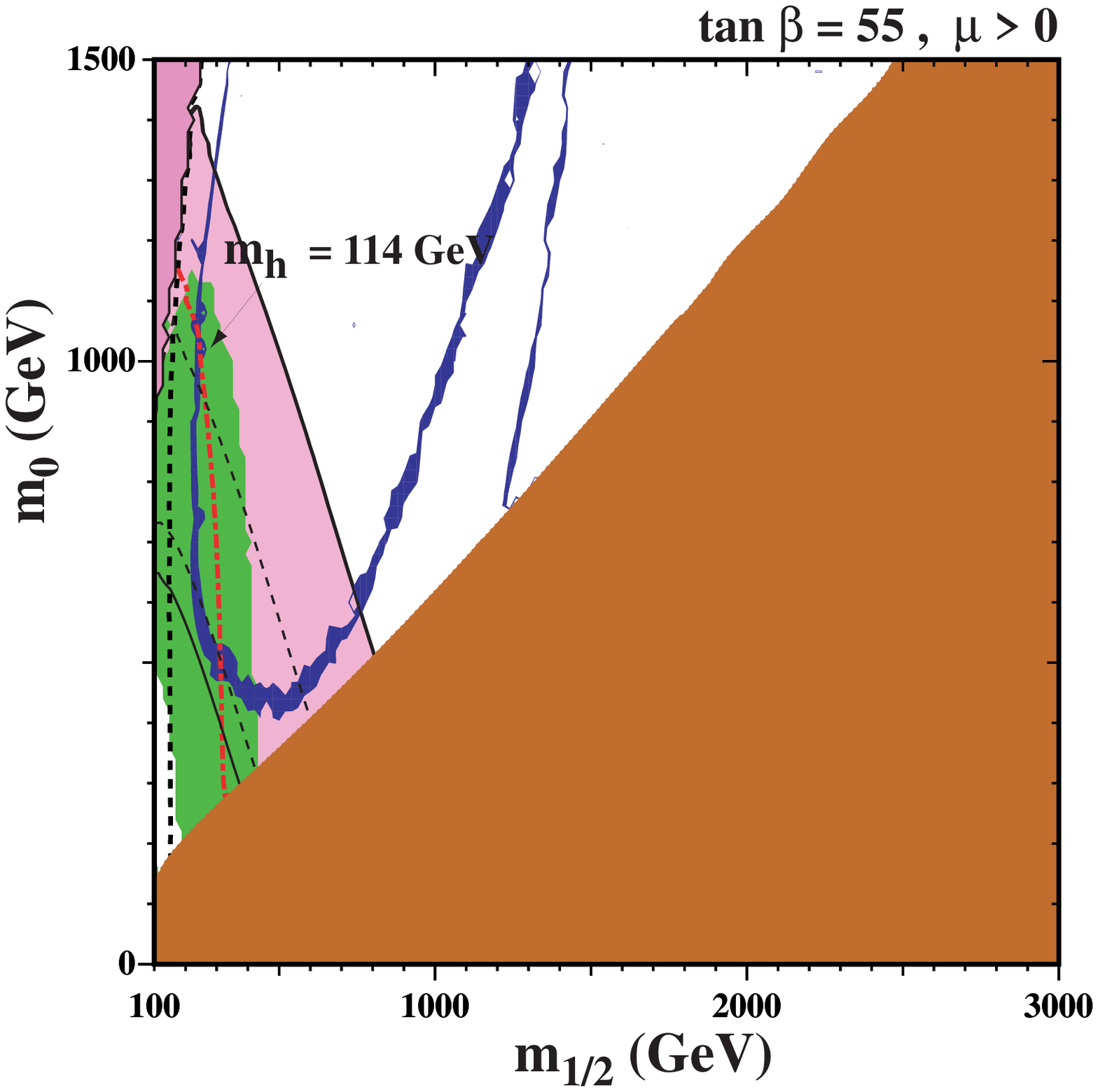}
  \caption{The $(m_{1/2}, m_0)$ planes for  (a) $\tan \beta = 10$ and  $\mu > 0$, 
assuming $A_0 = 0, m_t = 173.1$~GeV and
$m_b(m_b)^{\overline {MS}}_{SM} = 4.25$~GeV. The near-vertical (red)
dot-dashed lines are the contours $m_h = 114$~GeV, and the near-vertical (black) dashed
line is the contour $m_{\chi^\pm} = 104$~GeV.  The medium (dark
green) shaded region is excluded by $b \to s
\gamma$, and the dark (blue) shaded area is the cosmologically
preferred region. In the dark
(brick red) shaded region, the LSP is the charged ${\tilde \tau}_1$. The
region allowed by the E821 measurement of $g_\mu -2$ at the 2-$\sigma$
level, is shaded (pink) and bounded by solid black lines, with dashed
lines indicating the 1-$\sigma$ ranges. In (b), $\tan \beta= 55$.}
  \label{fig:UHM}
\end{figure}

For given values of $\tan \beta$, $A_0$,  and $sgn(\mu)$, the regions of the CMSSM
parameter space that yield an
acceptable relic density and satisfy other phenomenological constraints
may be displayed in the  $(m_{1/2}, m_0)$ plane.
In Fig. \ref{fig:UHM}a,  the dark (blue)
shaded region corresponds to that portion of the CMSSM plane
with $\tan \beta = 10$, $A_0 = 0$, and $\mu > 0$ such that the computed
relic density yields the WMAP value \cite{wmap} of 
\beq
\Omega h^2 = 0.111 \pm 0.006 .
\eeq
The bulk region at relatively low values of 
$m_{1/2}$ and $m_0$,  tapers off
as $m_{1/2}$ is increased.  At higher values of $m_0$,  annihilation cross sections
are too small to maintain an acceptable relic density and $\Omega_\chi h^2$ is too large.
At large $m_{1/2}$,
co-annihilation processes between the LSP and the next lightest sparticle 
(in this case the $\tilde \tau$) enhance the annihilation cross section and reduce the
relic density.  This occurs when the LSP and NLSP are nearly degenerate in mass.
The dark (red) shaded region has $m_{\tilde \tau}< m_\chi$
and is excluded.   The effect of coannihilations is
to create an allowed band about 25-50 GeV wide in $m_0$ for $m_{1/2} \la
950$ GeV, or $m_\chi \la 400$ GeV, which tracks above the $m_{{\tilde \tau}_1}
 =m_\chi$ contour~\cite{efo}.  Also shown in the figure are some phenomenological
 constraints from the lack of detection of charginos \cite{LEPsusy}, or Higgses
\cite{LEPHiggs} as well as constraints from $b \to s \gamma$ \cite{bsgex}
and $g_\mu - 2$ \cite{newBNL}.  The locations of these constraints are described in the caption.

At larger $m_{1/2}, m_0$ and $\tan \beta$, the relic neutralino
density may be reduced by rapid annihilation through direct-channel $H, A$ Higgs 
bosons, as seen in Fig.~\ref{fig:UHM}(b) \cite{funnel,efgosi}.
Finally, the relic density can again be brought
down into the WMAP range at large $m_0$ 
in the `focus-point' region close the boundary where EWSB 
ceases to be possible and the lightest neutralino $\chi$
acquires a significant higgsino component \cite{fp}. The start of the focus point
region is seen in the upper left of Fig.~\ref{fig:UHM}b. 

A global likelihood analysis enables one to 
pin down the available parameter space in the CMSSM. 
One can avoid the dependence on priors by performing
a pure $\chi^2$-based fit as 
done in~\cite{like2}
which used a Markov-Chain
Monte Carlo (MCMC) technique to explore efficiently the likelihood function in
the parameter space of the CMSSM. 

\begin{figure}[htb!]
  \includegraphics[width=0.6\textwidth]{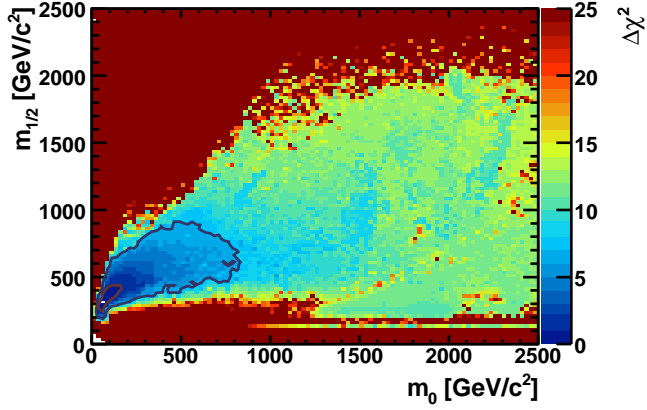}
  \caption{The $\Delta\chi^2$ function in the $(m_0, m_{1/2})$ plane for
  the CMSSM. We see that the coannihilation region at
  low $m_0$ and $m_{1/2}$ is favored.}
\label{fig:m0m12}       % Give a unique label
\end{figure}

The best fit point is shown in  Fig.~\ref{fig:m0m12},
which also displays contours of the $\Delta\chi^2$ function in the CMSSM
(solid curves outline the 68 and 95\% CL regions).
The parameters of the best-fit CMSSM point are
$m_{1/2} = 310 \gev$,  $m_0 = 60 \gev$,  $A_0 = 130 \gev$, and $\tb = 11$,
yielding the overall $\chi^2/{\rm N_{\rm dof}} = 20.6/19$ (36\% probability) 
and nominally $\Mh = 114.2 \gev$ \cite{like2}.
The best-fit point is in the coannihilation region of the
$(m_0, m_{1/2})$ plane. The C.L.\ contours extend to somewhat large
values of $m_0$.
However, the qualitative features of the $\Delta\chi^2$ contours 
indicate a preference for small
$m_0$ and $m_{1/2}$.

\begin{figure}[hbt]
  \includegraphics[width=0.5\textwidth]{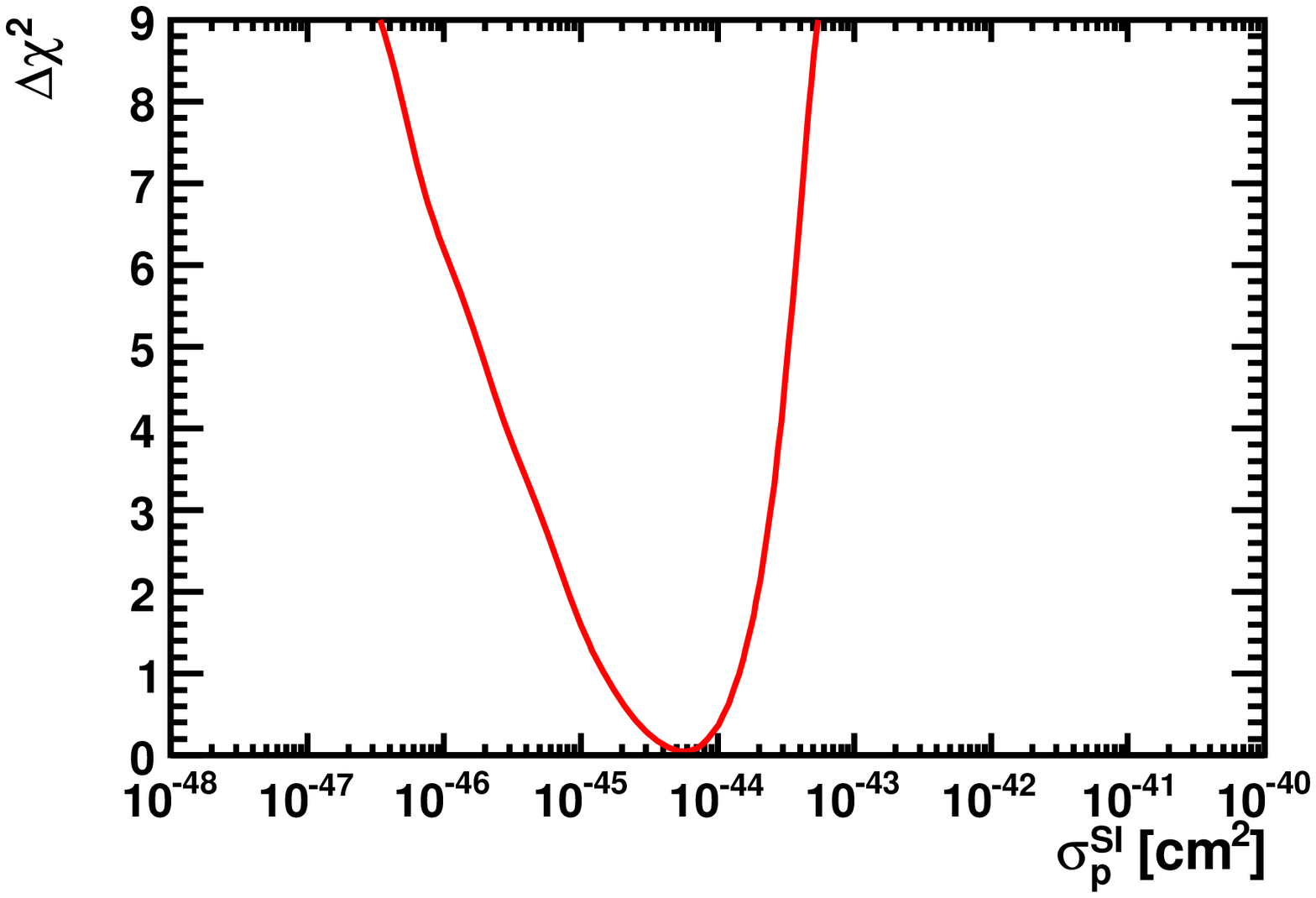}
  \includegraphics[width=0.5\textwidth]{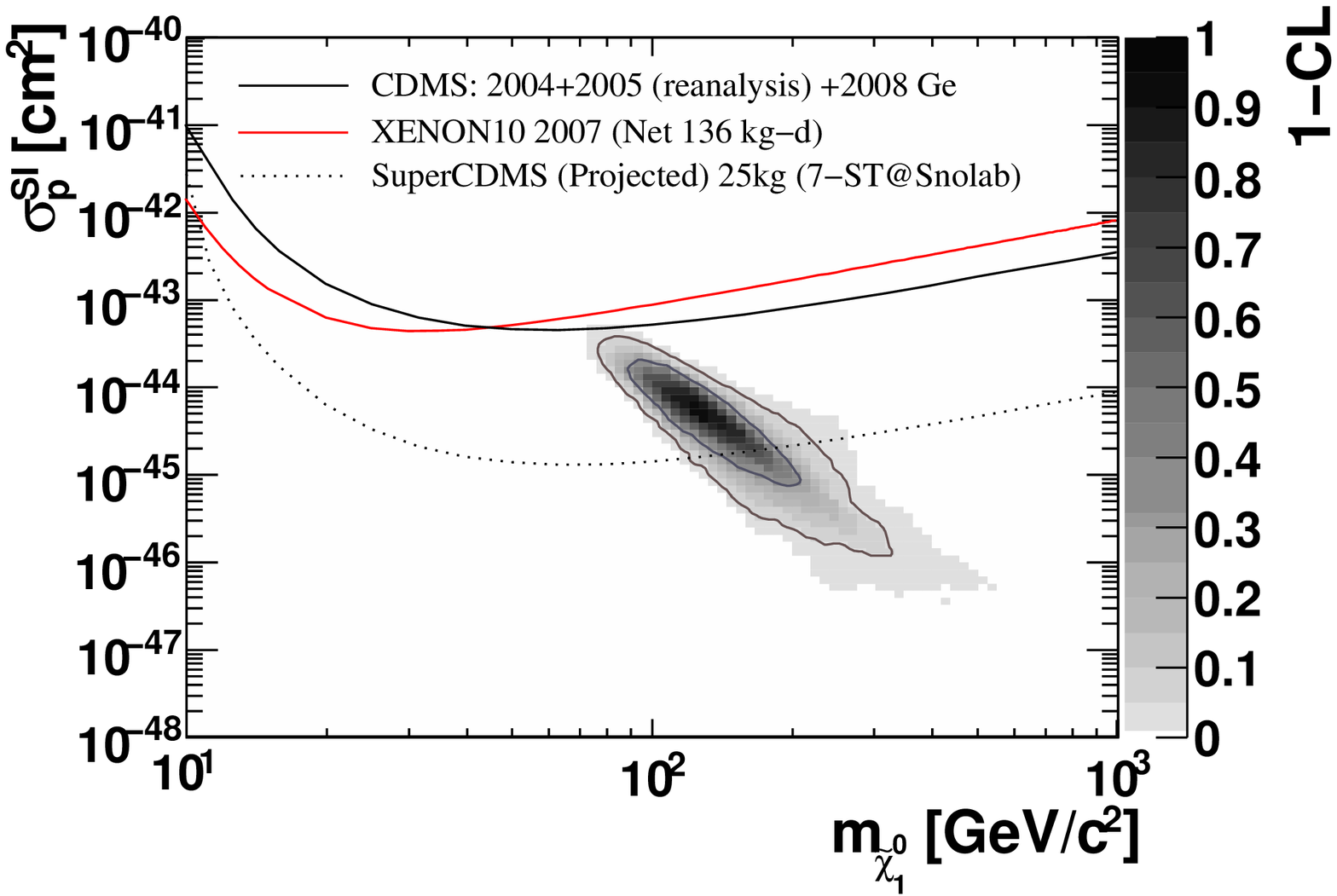}
\caption{The likelihood function for the
  spin-independent $\neu{1}$-proton scattering cross section $\ssi$
in the CMSSM (left panel). The correlation between the spin-independent DM
scattering cross section $\ssi$ and $\mneu{1}$
in the CMSSM (right panel).
\label{fig:sig}}
\end{figure}

 The frequentist analysis described above can also be used to predict
the neutralino-nucleon elastic scattering cross section \cite{like2}. 
The value of $\ssi$ shown in Fig.~\ref{fig:sig}a is calculated assuming a
  $\pi$-N scattering $\sigma$ term 
$\Sigma_{\pi N} = 64$~MeV.
We see in Fig.~\ref{fig:sig} that values of the $\neu{1}$-proton cross
section $\ssi \sim 10^{-8}$~pb are expected in the CMSSM, and that much
larger values seem quite unlikely. The 2D $\chi^2$ function in the 
($m_\chi, \sigma_p$) plane is shown in Fig. ~\ref{fig:sig}b.

\section{mSUGRA Models}

As noted above, mSUGRA models are a very constrained version of the MSSM,
as $\tan \beta$ is no longer a free parameter.
In minimal $N=1$ supergravity, the K\"ahler
potential can be written as
\beq
G = K(\phi^i,{\phi_i}^*,z,z^*) +  \ln ( \kappa^6 |W|^2)
\label{min}
\eeq
with 
\beq
K = \kappa^2( \phi^i {\phi_i}^*  + z z^*) 
\eeq
where $W = f(z) + g(\phi)$ is the superpotential, 
assumed to be separable in hidden sector fields, $z$, and
standard model fields, $\phi$. $\kappa^{-1} = M_P/\sqrt{8\pi}$ and the
Planck mass is $M_P = 1.2\times 10^{19}$ GeV. The scalar potential can be derived once
the superpotential is specified.  The simplest choice for a single hidden sector field is
a superpotential  \cite{Polonyi} $f(z) = m ( z + b)$. Using this in Eq. (\ref{min})
and dropping terms inversely proportional to the Planck mass, we can write \cite{bfs}
\begin{eqnarray}
V  & =  &  |{\partial g \over \partial \phi}|^2 +
m_{3/2} (\phi {\partial g \over \partial \phi} - \sqrt{3} g
+ h.c.) )  + m_{3/2}^2 \phi \phi^*  ,
\label{pot}
\end{eqnarray}
where the vacuum expectation value $\langle \kappa z \rangle = \sqrt{3} -1$ (with $\kappa b  = 2 - \sqrt{3}$,
chosen to cancel the vacuum energy density at the minimum) has been inserted and the
superpotential has been rescaled by a factor $e^{- \kappa b}$.

The first term in Eq. (\ref{pot}) is
the ordinary $F$-term part of the scalar potential of global supersymmetry. 
The next term, proportional to $m_{3/2}$ represents a universal trilinear
$A$-term. This can be seen by noting that $\sum \phi \partial g / \partial
\phi = 3 g$, so that in this model of supersymmetry breaking, $A_0 = (3 -
\sqrt{3}) m_{3/2}$.  Note that if the superpotential contains bilinear
terms, we would find $B_0 = (2 - \sqrt{3}) m_{3/2}$. The last term represents a
universal scalar mass of the type advocated in the CMSSM, with
$m_0^2 = m_{3/2}^2$. The generation of such soft terms is a rather generic
property of low energy supergravity models \cite{mark} and the relation 
$B_0 = A_0 - m_0$ is derived from the minimal form of the K\"ahler potential
and does not depend on the specific form of $f(z)$. 

\begin{figure}[htb!]
  \includegraphics[width=.45\textwidth]{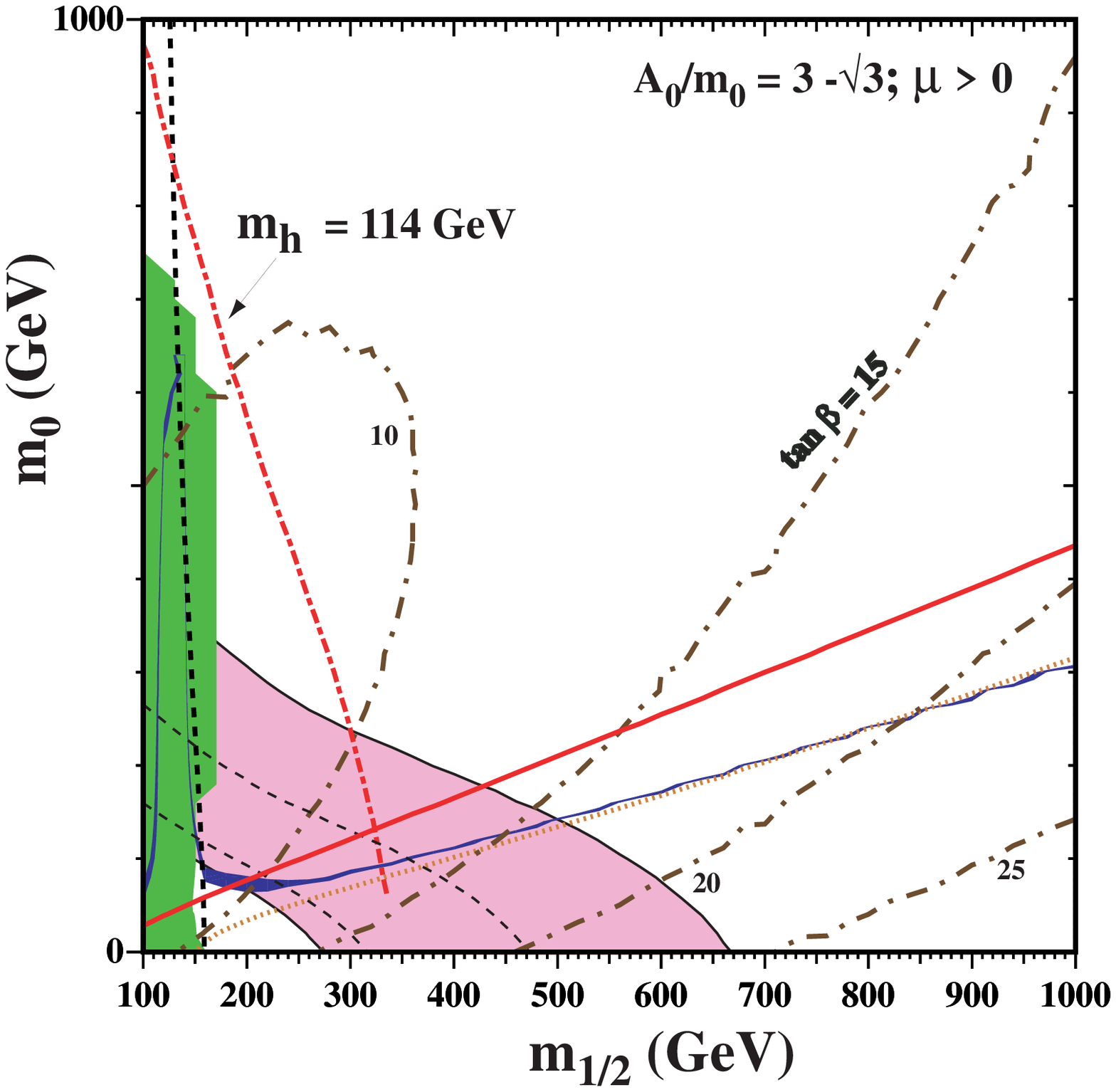}
    \includegraphics[width=.45\textwidth]{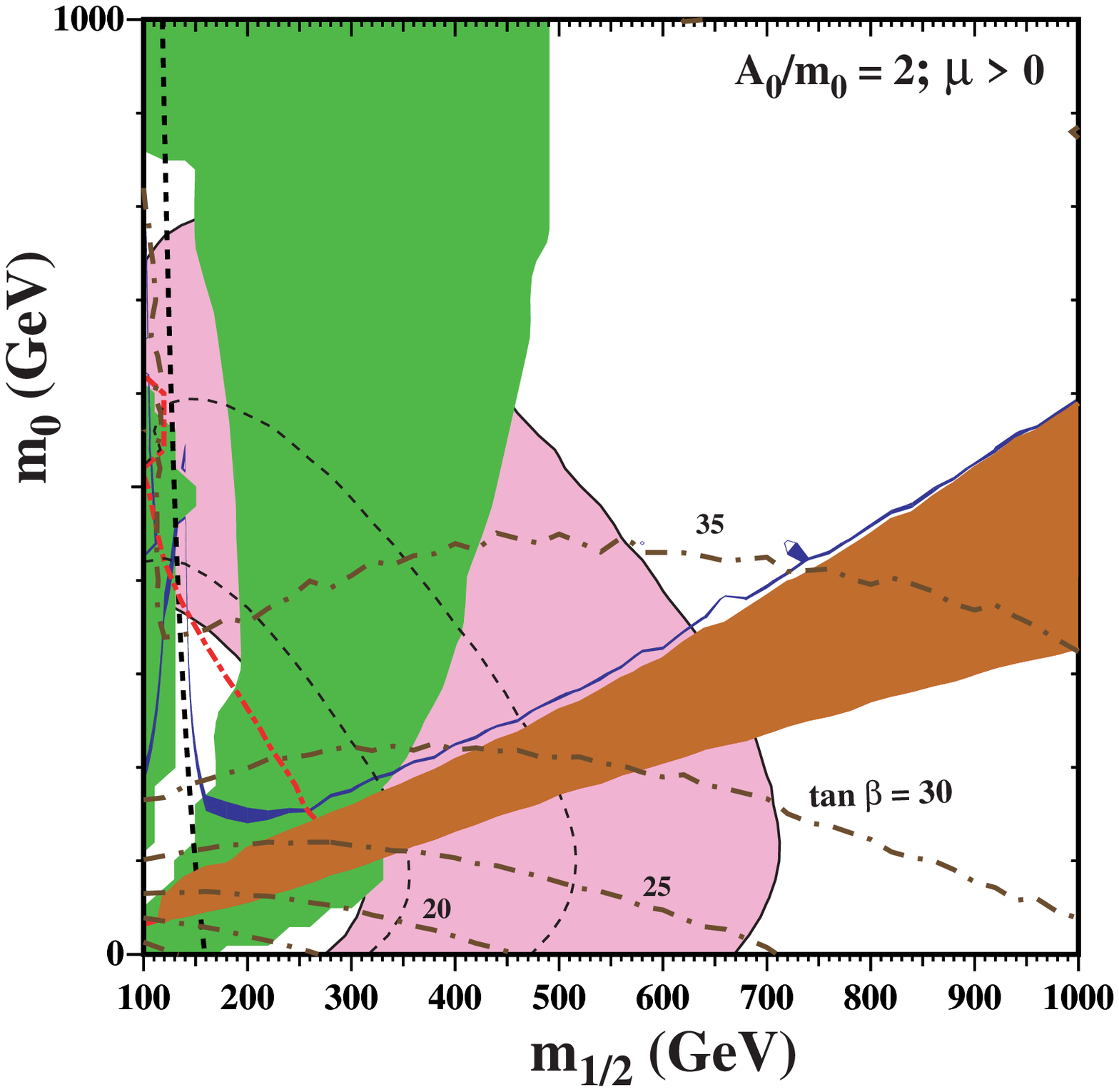}
  \caption{As in Fig. \ref{fig:UHM}, the $(m_{1/2}, m_0)$ planes for  (a) $A_0/m_0 =3 - \sqrt{3}$,
  $B_0 = A_0 - m_0$,
  and  $\mu > 0$. The dot-dashed contours show curves of constant $\tan \beta$. 
  Below the dark (brown) line the gravitino is the LSP. 
In (b), $A_0/m_0 = 2$. }
  \label{fig:msug}
\end{figure}

The analogue of the CMSSM ($m_{1/2},m_0$) plane for mSUGRA models is shown in Fig.
\ref{fig:msug} for two fixed values of $A_0/m_0$ \cite{vcmssm}. In the left panel, the Polonyi model
choice of $A_0/m_0 = 3 - \sqrt{3}$ is made.  As one can see, each point on the plane
corresponds to a specific value of $\tan \beta$ as specified by the dot-dashed contours.
Also seen in this panel is a solid (brown) diagonal line. Below this line, 
the gravitino is the LSP, since $m_\chi \approx 0.43 m_{1/2} < m_0 = m_{3/2}$.
Below this line, there is a diagonal dotted (red) line, which separates
the region for which the next lightest supersymmetric particle is 
the neutralino (above) or the stau (below).  The area below this
line is not shaded since in principle, this area could be allowed as
the stau is not stable.  However, constraints from big bang nucleosynthesis
excluded much of gravitino LSP region in mSUGRA models \cite{cefos}.

In the right panel of Fig.~\ref{fig:msug}, $A_0/m_0 = 2$.  In this case, the boundary
of $m_{\tilde \tau} = m_\chi$ is above the boundary for $m_\chi = m_{3/2}$. 
As a consequence, there is now a viable co-annihilation strip with neutralino dark matter.
Below that, there is an excluded region (shaded dark red) which has a stau LSP. 
Below the shaded region, the gravitino is again the LSP subject to BBN constraints \cite{cefos}.
Funnel and focus point regions are generally absent in mSUGRA models.

\section{No-Scale Models}
\begin{figure}[htb!]
  \includegraphics[width=.80\textwidth]{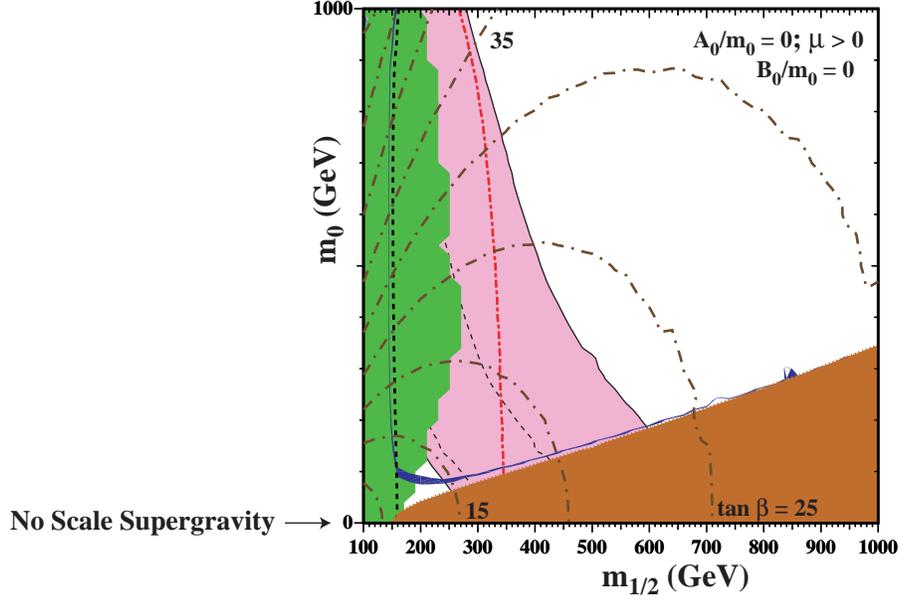}
  \caption{As in Fig. \ref{fig:UHM}, the $(m_{1/2}, m_0)$ planes for  (a) $A_0/m_0 =
  B_0/m_0 = 0$,  
  and  $\mu > 0$. The dot-dashed contours show curves of constant $\tan \beta$.  }
  \label{fig:noscale}
\end{figure}

No-scale supergravity models \cite{nosc1} are characterized by
a K\"ahler potential defined by
\beq
K = -3 \ln \left[ \kappa(z + z^*) - \frac{1}{3}\kappa^2 \phi^i \phi_i^* \right] .
\eeq
The scalar potential takes the simple form 
\beq
V = e^{G - \frac{1}{3}K}\left| \frac{\partial g}{\partial \phi^i} \right|^2 .
\eeq
Thus one immediately finds that $m_0 = A_0 = B _0 = 0$, and the only source of
supersymmetry breaking is transmitted through the gaugino masses
(if the gauge kinetic function is a non-trivial function of $z$).
In Fig. \ref{fig:noscale}, an ($m_{1/2}, m_0$) plane with $A_0/m_0 = B_0/m_0 = 0$ is shown.
Like the mSUGRA model, specification of $B_0$ fixes $\tan \beta$ at each point on the plane.
However, unlike mSUGRA models, the gravitino mass is rather arbitrary and can be set 
independently from other supersymmetry-breaking scales.
No-scale models run along the $m_0 = 0$ axis of this plane.
As one can see, along $m_0 = 0$, there is no way to achieve a sufficiently high relic
density to match the WMAP determination and for $m_{1/2} \la 150$ GeV, there
is a problem with the branching ratio for $B_s \to X_s \gamma$, while at larger
$m_{1/2}$, the LSP is either the stau or the gravitino.  Furthermore, the LEP limit on the Higgs 
mass requires $m_{1/2} \ga 340$ GeV. Thus it would appear that no-scale models
are not phenomenologically viable.

\section{SuperGUT Models}

In all of the models discussed above, it was assumed that the renormalization
scale at which scalar and gaugino masses are unified, $M_{in}$, is the same scale 
at which gauge coupling unification occurs, $M_{GUT}$.  This need not be the case.
Indeed, since supersymmetry breaking may occur either below or above $M_{GUT}$,
it is quite possible that $M_{in} \ne M_{GUT}$.  
For example, as $M_{in}$ is decreased below $M_{GUT}$, the differences between the
renormalized sparticle masses diminish and the regions of the $(m_{1/2}, m_0)$
planes that yield the appropriate density of cold dark matter move away from the boundaries~\cite{EOS06}.
Eventually, for small $M_{in}$, the coannihilation and focus-point regions of the conventional
GUT-scale CMSSM merge and for very small $M_{in}$ they disappear 
entirely.

On the other hand,  increasing
$M_{in}$ increases the renormalization of the sparticle masses which tends to increase
the splittings between the physical sparticle masses~\cite{pp}. Furthermore, 
this in turn has the effect of increasing the relic density in much of the $(m_{1/2}, m_0)$
plane. As a consequence, the coannihilation strip is squeezed to lower values of $m_{1/2}$
\cite{Calibbi,emo},
particularly for $\tan \beta \sim 10$, and
even disappears as $M_{in}$ increases. At the same time, the focus-point strip often moves out
to ever larger values of $m_0$.
The allowed region of parameter space that survives longest is the rapid-annihilation
funnel at large $m_{1/2}$ and $\tan \beta$. In the CMSSM with $M_{in} = M_{GUT}$, 
the funnel region also requires large
$m_0$ and would make a contribution to $g_\mu - 2$ that is
too small to explain the experimental discrepancy with Standard Model calculations
based on low-energy $e^+ e^-$ data. However,  for large $M_{in}$,
the funnel region extends to low $m_0$ (including $m_0 = 0$) and in some cases
will be compatible with the $g_\mu - 2$ measurements.

For superGUT models with $M_{in} > M_{GUT}$, one must specify in addition to 
Eq. (\ref{suppot}), the GUT superpotential which for minimal SU(5) takes the form, 
\begin{eqnarray}
W_5 &=& \mu_\Sigma Tr \hat{\Sigma}^2 + \frac{1}{6}\lambda' Tr\hat{\Sigma}^3
 + \mu_H \hat{\calh}_{1\alpha} \hat{\calh}_2^{\alpha} 
 + \lambda \hat{\calh}_{1\alpha}\hat{\Sigma}^{\alpha}_{\beta} \hat{\calh}_2^{\beta} \nl
 +({\bf h_{10}})_{ij} \epsilon_{\alpha\beta\gamma\delta\zeta}
   \hat{\psi}^{\alpha\beta}_i \hat{\psi}^{\gamma\delta}_j \hat{\calh}_2^{\zeta} 
 +({\bf h_{\overline{5}}})_{ij} \hat{\psi}^{\alpha\beta}_i \hat{\phi}_{j\alpha} \hat{\calh}_{1\beta} ,
\label{W5}
\end{eqnarray}
where the $d^c$ and $L$ superfields of the MSSM reside in the $\bf{\overline{5}}$
representation, $\hat{\phi}_i$, while the $Q,\ u^c$ and $e^c$ superfields are in the $\bf{10}$ representation,
$\hat{\psi}_i$. 
$\hat{\Sigma}(\bf{24})$ is the SU(5) adjoint Higgs
multiplet and the two Higgs doublets of the MSSM, ${H}_1$ and ${H}_2$
are extended to five-dimensional SU(5) representations $\hat{\calh}_1(\bf{\overline{5}})$ and 
$\hat{\calh}_2(\bf{5})$ respectively.
There are now two new couplings: $\lambda$ and $\lambda'$.  $\lambda$ affects
directly the running of the soft Higgs masses, adjoint and Yukawa couplings, while
$\lambda'$ affects only the adjoint. Accordingly there are also new soft masses and $\mu$
terms associated with SU(5).

A superGUT version of the CMSSM based on SU(5) is now a 7+ parameter theory \cite{emo}
specified by $m_{1/2}, m_0, A_0$, $\tan \beta$, and $sgn(\mu)$ as in the CMSSM, as well
as $M_{in}, \lambda$, and $\lambda'$. At $M_{in} > M_{GUT}$, the universality conditions
become, 
\begin{eqnarray}
\mfivl=\mtenl=\mfiv=\mten=m_{\calh_1}=m_{\calh_2}=m_{\Sigma} &\equiv & m_0, \nnl
\afiv=\aten=A_{\lambda}=A_{\lambda'} &\equiv & A_0, \nnl
M_5 &\equiv & m_{1/2}.
\end{eqnarray}
These parameters are evolved down to $M_{GUT}$ and matched to their Standard Model
CMSSM counterparts.  

\begin{figure}[htb!]
  \includegraphics[width=.45\textwidth]{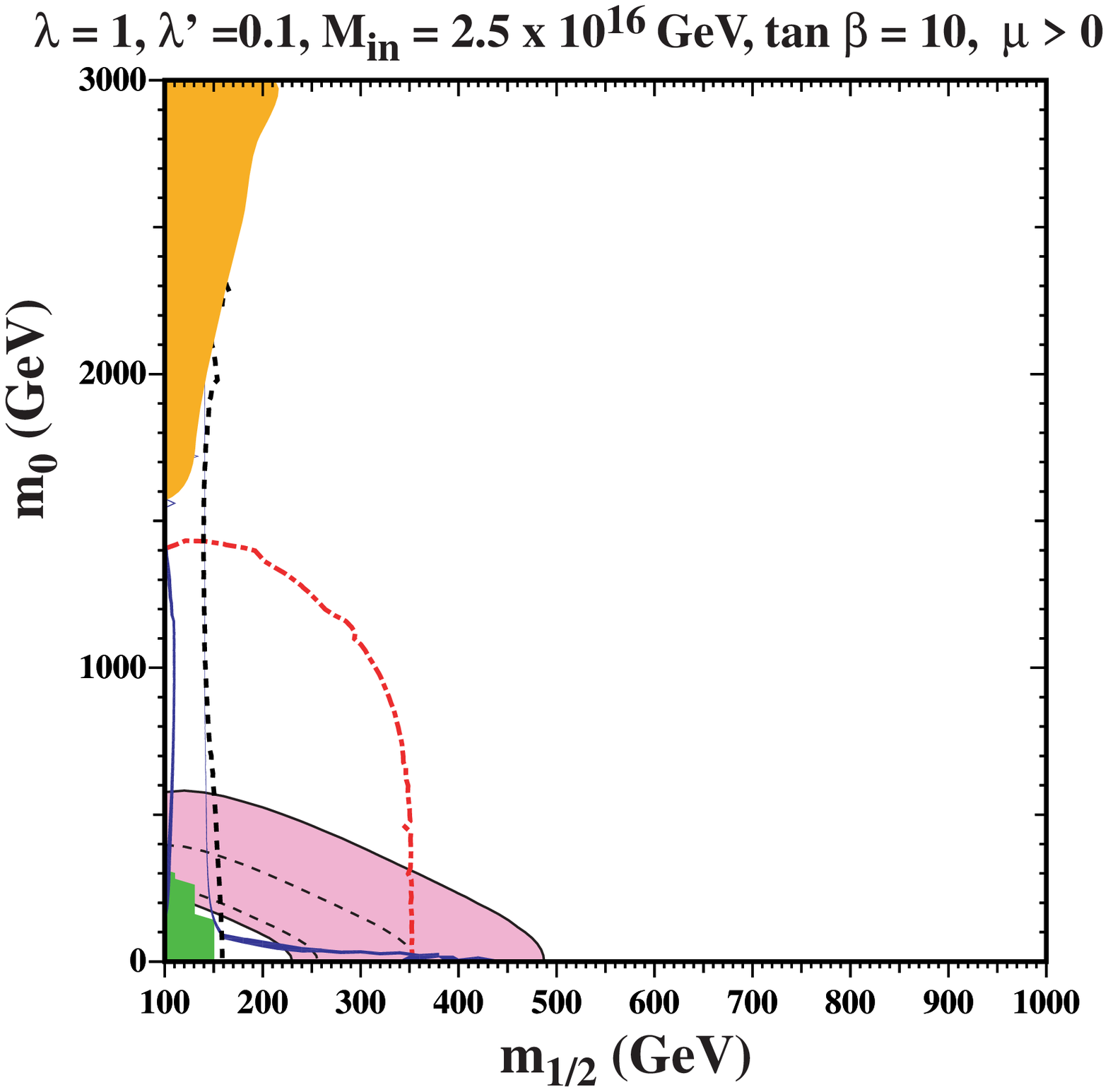}
    \includegraphics[width=.45\textwidth]{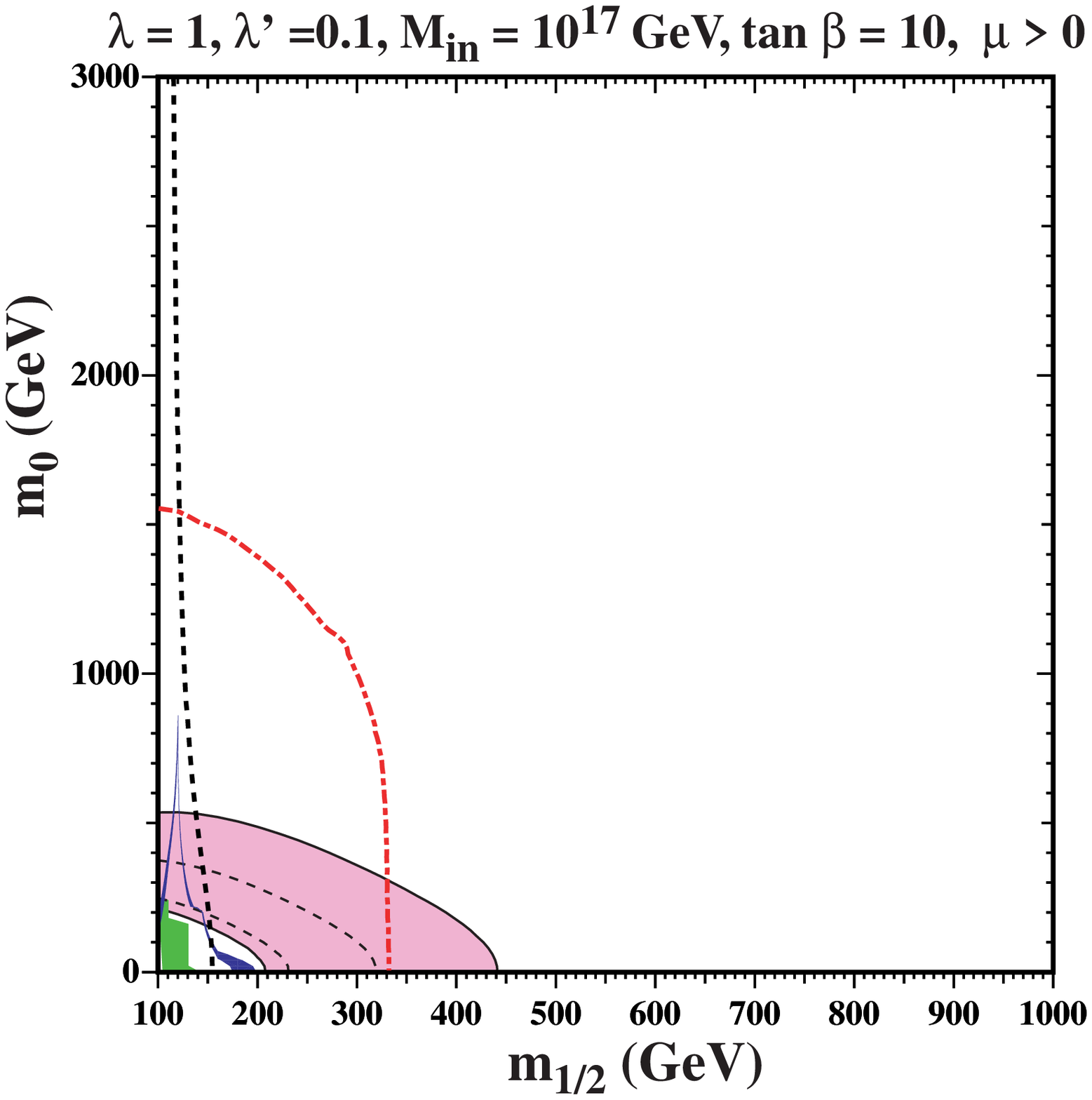}
  \caption{As in Fig. \ref{fig:UHM}, the $(m_{1/2}, m_0)$ planes for  
  $A_0 = 0, \tan \beta = 10, \lambda = 1$
and $\lambda' = 0.1$ for different choices of $M_{in}$:
(a) $2.5 \times 10^{16}$~GeV  and (b) $10^{17}$~GeV. 
 }
  \label{sg10}
\end{figure}

Some examples of ($m_{1/2}, m_0$) planes in a superGUT model are shown in Fig. \ref{sg10}
for $\tan \beta = 10$
and two values of $M_{in}$ and the specific choices $\lambda = 1, \lambda' = 0.1$ \cite{emo}.
These results are quite
insensitive to the value of $\lambda'$ and can be compared with the CMSSM plane shown in 
Fig.\ref{fig:UHM}a.
For $M_{in} = 2.5 \times 10^{16}$~GeV,  we see two dramatic effects from the modest
increase in $M_{in}$. One is the rapid disappearance of the stau LSP region, which
has retreated to $m_0^2 < 0$.
In the particular example shown, the coannihilation strip extends to $m_{1/2} \sim 450$~GeV,
and there is a healthy portion compatible with the $g_\mu - 2$ constraint. Note that the
chargino, $m_h$, $g_\mu - 2$ and $b \to s \gamma$
constraints are relatively stable in the $(m_{1/2}, m_0)$ plane with respect to changes in $M_{in}$.
The other noticeable feature in 
panel (a) of Fig.~\ref{sg10} is the retreat of the EWSB
constraint to smaller $m_{1/2}$ and larger $m_0$. This effect
is quite sensitive to the value of $\lambda$, whereas the fate of the coannihilation region is
relatively insensitive to its value.

For the choice $M_{in} = 10^{17}$~GeV, shown in panel  (b) of Fig.~\ref{sg10},
these effects are more pronounced: both the coannihilation and the focus-point strips
have disappeared entirely. There is a small piece of the $(m_{1/2}, m_0)$ plane where
the relic density falls within the WMAP range, but this is incompatible with $m_h$ and
gives too large a value of $g_\mu - 2$. 

\begin{figure}[htb!]
  \includegraphics[width=.45\textwidth]{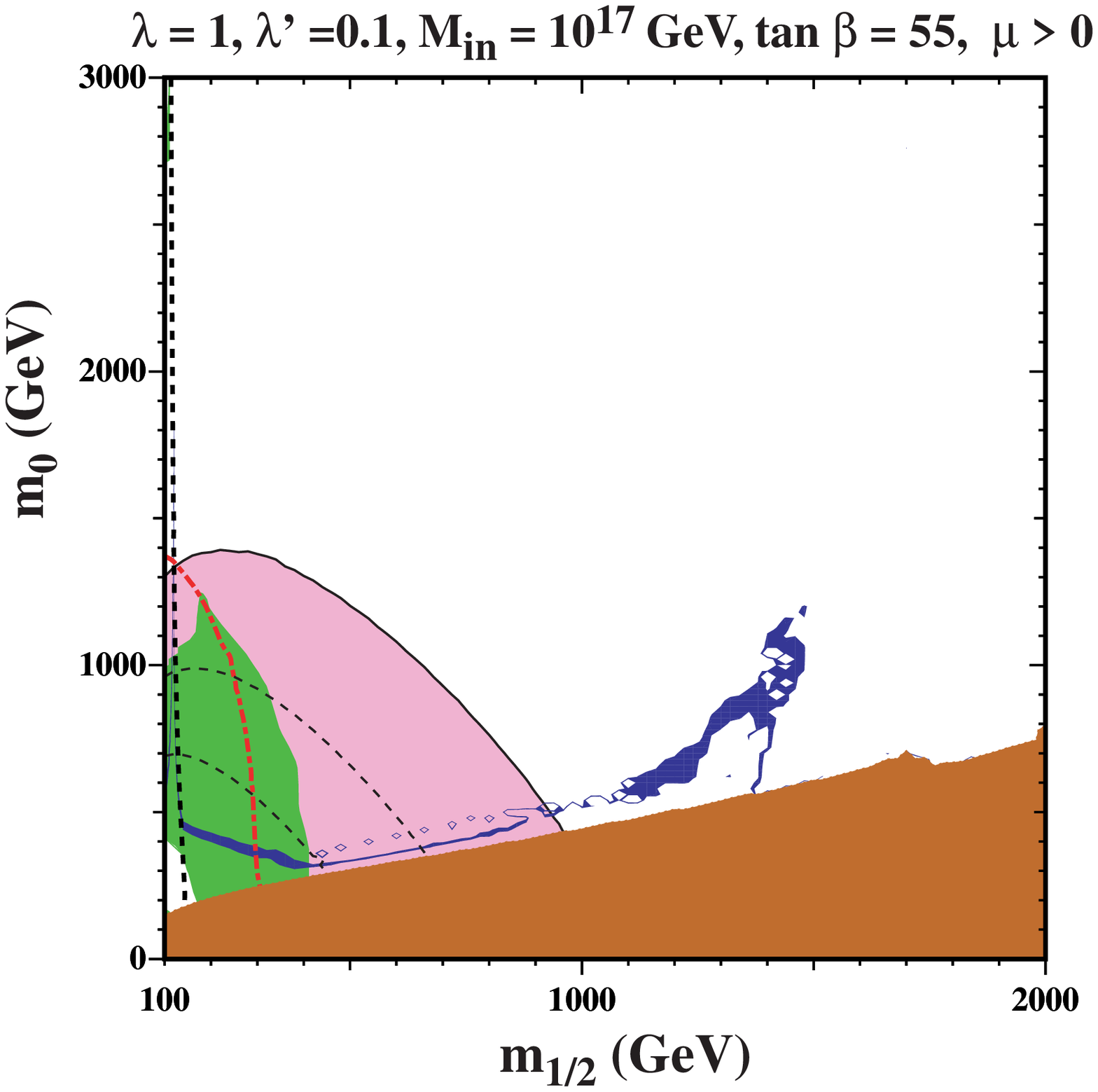}
    \includegraphics[width=.45\textwidth]{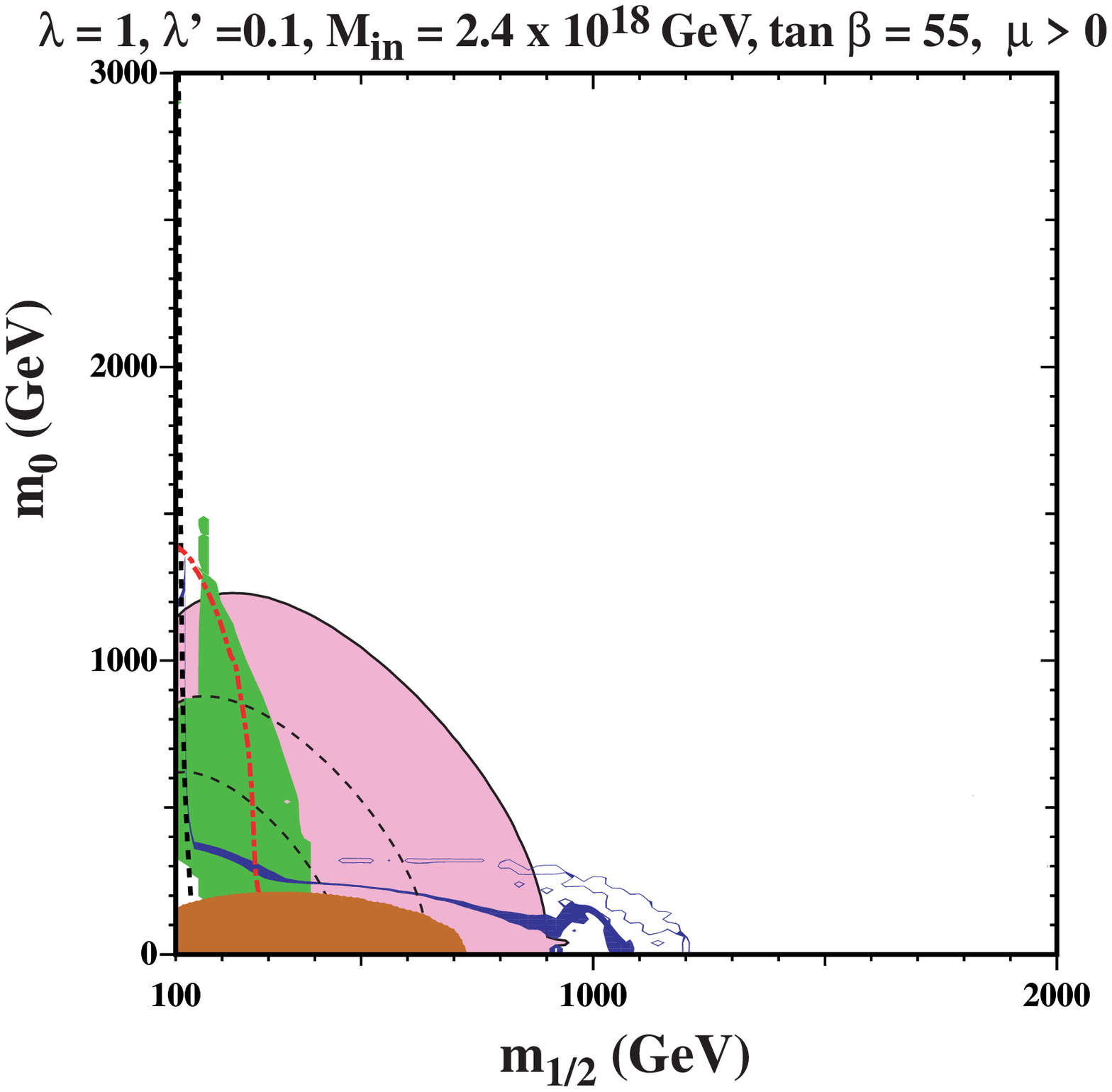}
  \caption{As in Fig. \ref{sg10},  for  
  $A_0 = 0, \tan \beta = 55, \lambda = 1$
and $\lambda' = 0.1$ for different choices of $M_{in}$:
(a) $10^{17}$~GeV  and (b) $2.4 \times 10^{18}$~GeV. 
 }
  \label{sg55}
\end{figure}

For $\tan \beta = 55$, we see in Fig. \ref{sg55} that as 
$M_{in}$ increases, the renormalization effects  
cause the stau LSP region to retreat as in the $\tan \beta = 10$
case, though more slowly, and it does not disappear entirely, even for
$M_{in} = 2.4 \times 10^{18}$~GeV. Likewise, whilst the coannihilation strip shrinks
with increasing $M_{in}$, it does not disappear, and much of it remains consistent with
$m_h$, $b \to s \gamma$ and $g_\mu - 2$. The rapid-annihilation funnel also persists
as $M_{in}$ increases, staying in a similar range of $m_{1/2}$, but
shifting gradually to lower values of $m_0$. In particular, we note that for the case
$M_{in} =  2.4 \times 10^{18}$~GeV, the no-scale possibility 
$m_0 = 0$~\cite{nosc1,eno} is allowed, on one or both flanks of the rapid-annihilation funnel. 
Finally, we note that the
EWSB boundary disappears entirely for the displayed choices of
$M_{in} > M_{GUT}$, as does the focus-point WMAP strip.

\section{No-Scale Models Resurrected}

As discussed above, no-scale models with GUT scale universality conditions
are not phenomenologically viable.  However, we have also seen when
the universality scale in increased above the GUT scale, no-scale
requirements such as $m_0 = 0$ may yield satisfactory models \cite{eno,SS,emo2}.

A no-scale superGUT model could be constructed from the same 
superpotential (\ref{W5}) with the additional requirements that
$m_0 = A_0 = B_0 = 0$ at $M_{in}$.  The latter condition translates to
$B_\Sigma = B_H = B_0 = 0$ at $M_{in}$. And it would appear therefore that we
now have a 4+ parameter theory specified by $m_{1/2}, M_{in}, \lambda, \lambda'$, and $sgn(\mu)$.
However, the model really only depends on the ratio of the two Higgs couplings, $\lambda/\lambda'$
and we are in fact left with a 3+ parameter theory.  As in the mSUGRA models discussed above,
$\tan \beta$ is determined from the EWSB conditions 
though the boundary condition is now $B_0 = 0$.  Since this condition is applied at $M_{in}$,
the MSSM (subGUT) $B$ parameter must be matched to the GUT $B$ parameters, $B_\Sigma$
and $B_H$ \cite{Borzumati},
\beq
B=B_H-\frac{6\lambda}{\mu\lambda'}\left[ (B_\Sigma -A_\lambda')(2B_\Sigma -A_\lambda')+m_\Sigma^2 \right] .
\label{Bmatch}
\eeq

\begin{figure*}[htb!]
      \includegraphics[width=.45\textwidth]{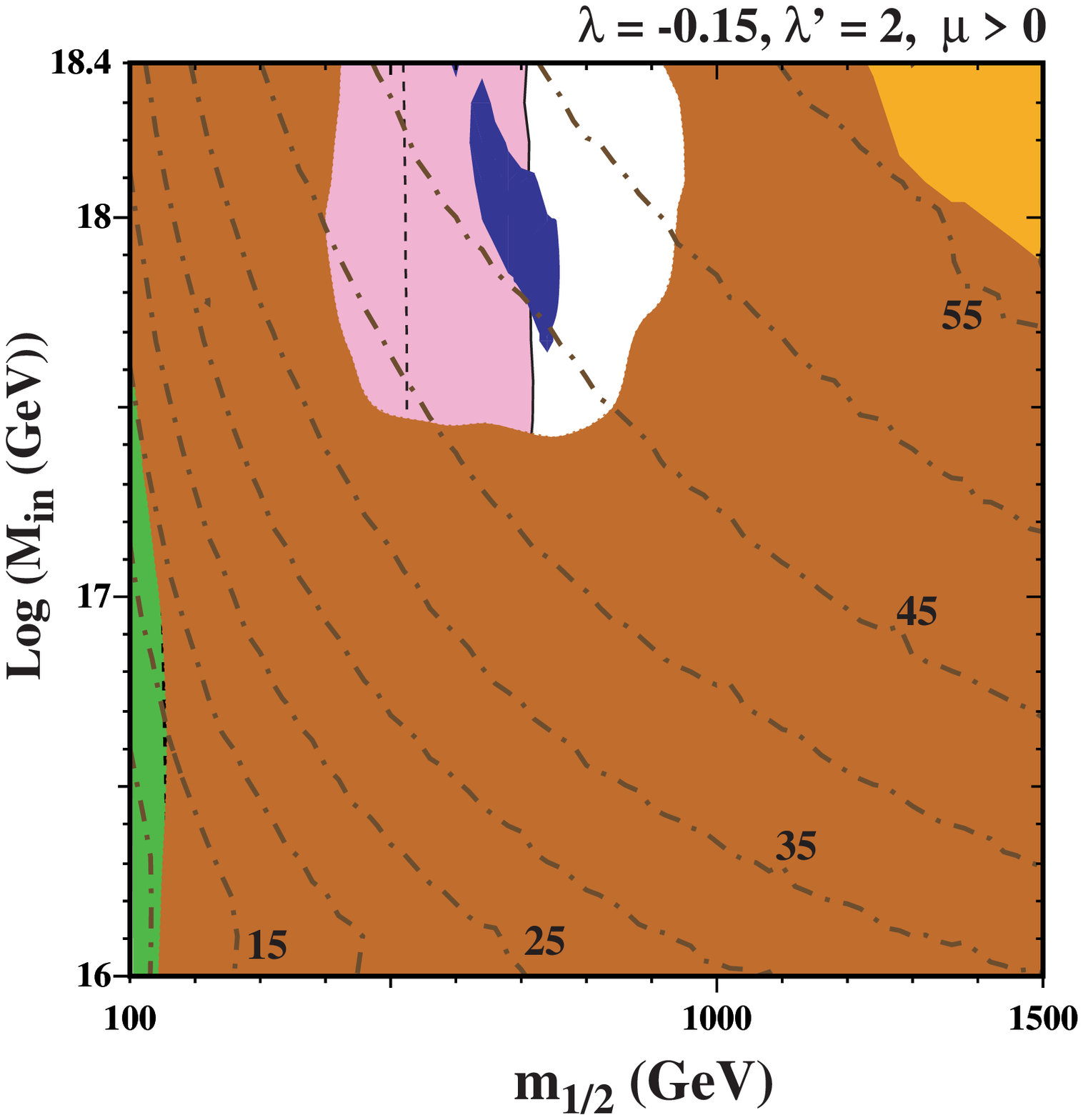}
    \includegraphics[width=.45\textwidth]{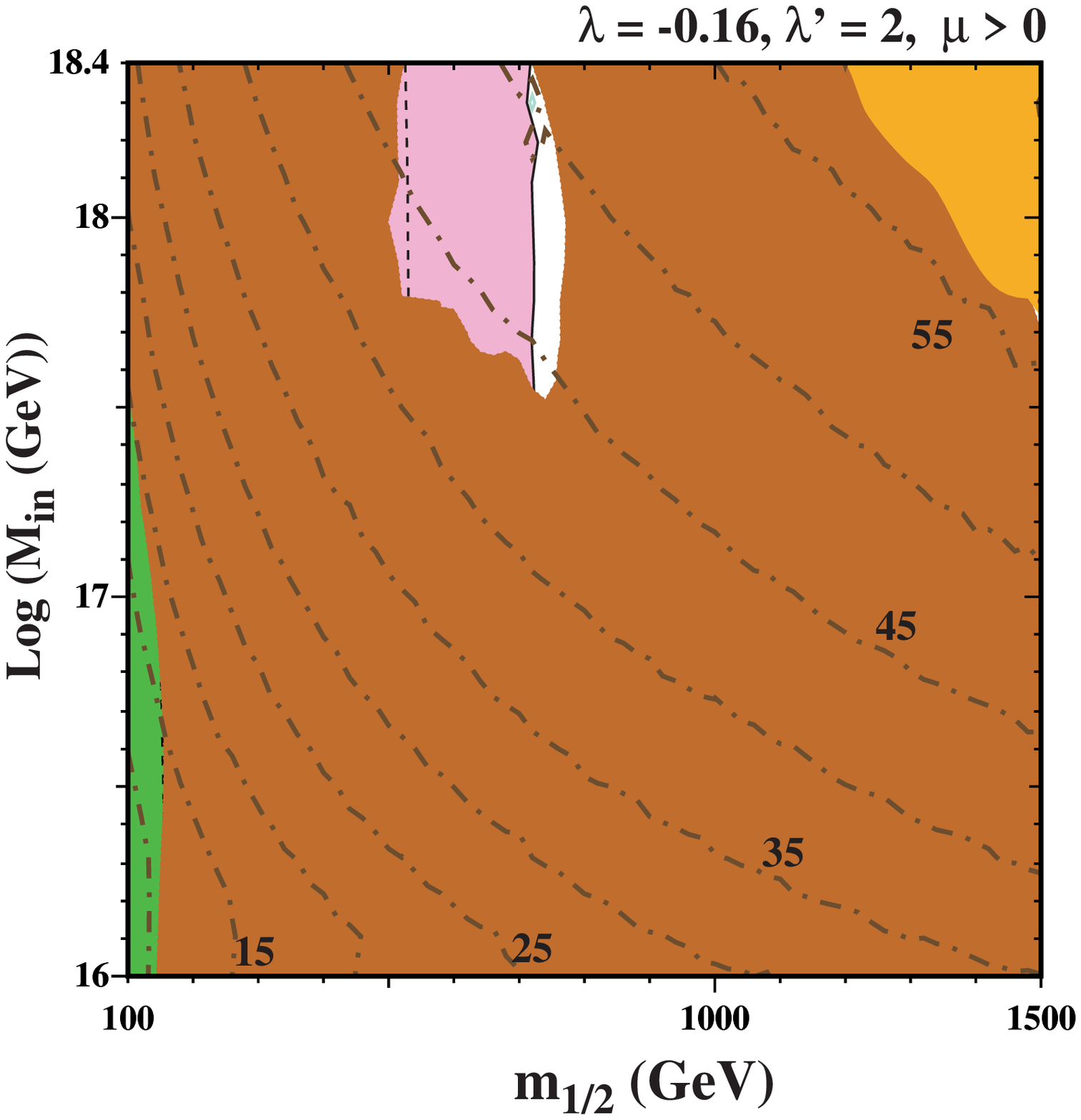}
  \caption{The $(m_{1/2}, M_{in})$ planes for the no-scale supergravity model with $\lambda' = 2$ and
(a) $\lambda = 0$, (b) $\lambda = -0.14$, (c) $\lambda = -0.15$ and (d)
$\lambda = -0.16$. Shading and line types are as in Fig. \ref{fig:UHM} with the 
exception that in the orange region we find no consistent solutions
to the RGEs. 
 }
  \label{noscsgut}
\end{figure*}

In Fig. \ref{noscsgut}, some examples of ($m_{1/2}, M_{in}$) planes are shown for
fixed $\lambda' = 2$ with $\lambda = 0, -0.14, -0.15$, and -0.16 \cite{emo2}.
 We see that in the first  panel of Fig.~\ref{noscsgut} the WMAP-compatible region
takes the form of a thin L-shaped strip in the $(m_{1/2}, M_{in})$ plane with a rounded corner:
points above and to the right of the L have values of $\ohsq$ that are too large.
The near-horizontal part of the line is located at $M_{in} \sim 5 \times 10^{16}$~GeV and 
extends from $m_{1/2} \sim 200$~GeV to $\sim 1000$~GeV, larger values being
excluded by the requirement that the LSP not be charged, and 
we find that $\tan \beta \in (16, 30)$. 
All the base strip is compatible with the LEP chargino
constraint, and with $b \to s \gamma$. However, only the portion with $m_{1/2} \ga
300$~GeV is compatible with the LEP lower limit on $m_h$. 
The near-vertical part of the no-scale strip is always
incompatible with the LEP Higgs constraint and (mostly) $b \to s \gamma$. 

\begin{figure}[htb!]
  \includegraphics[width=.44\textwidth]{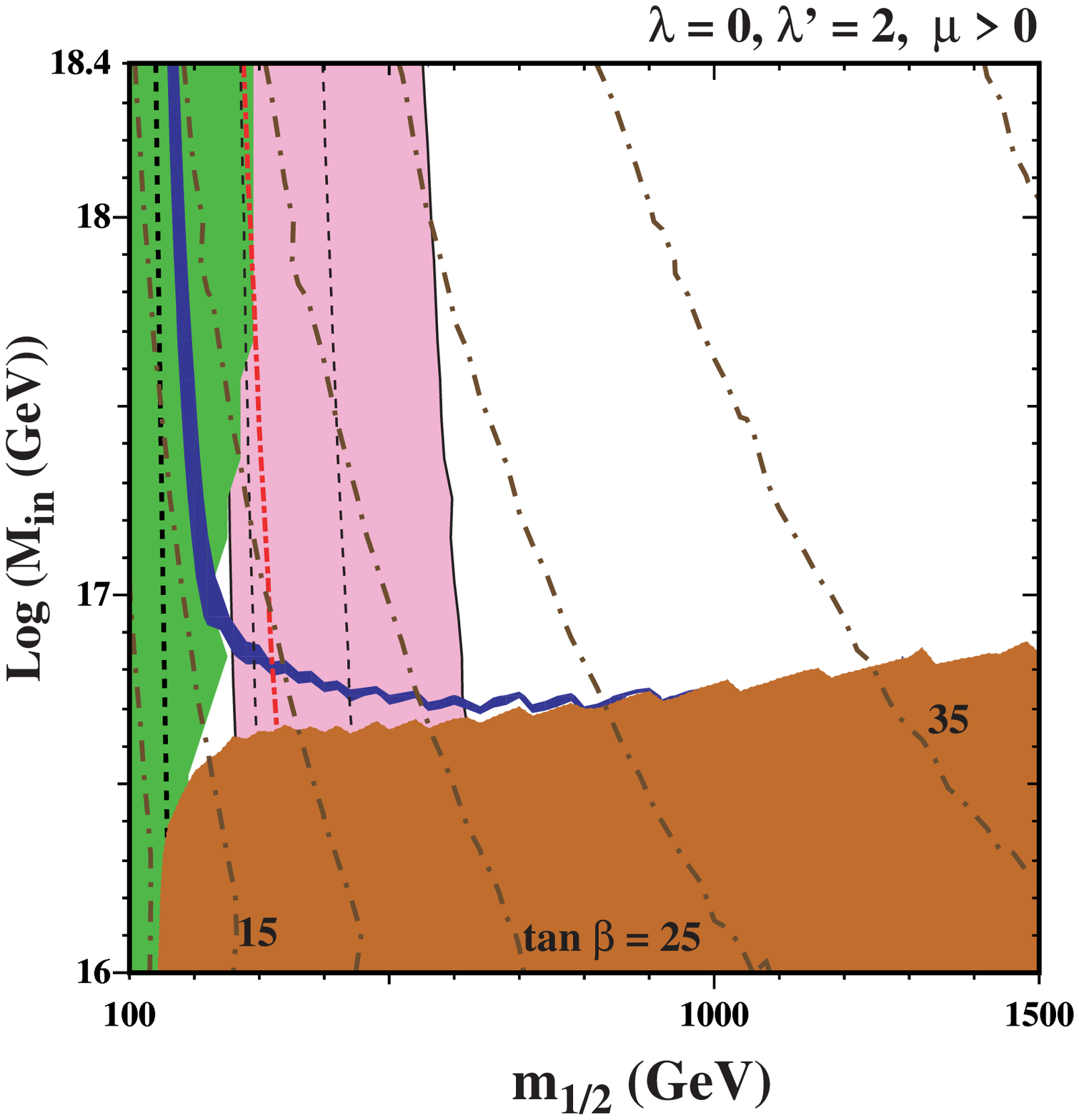}
    \includegraphics[width=.45\textwidth]{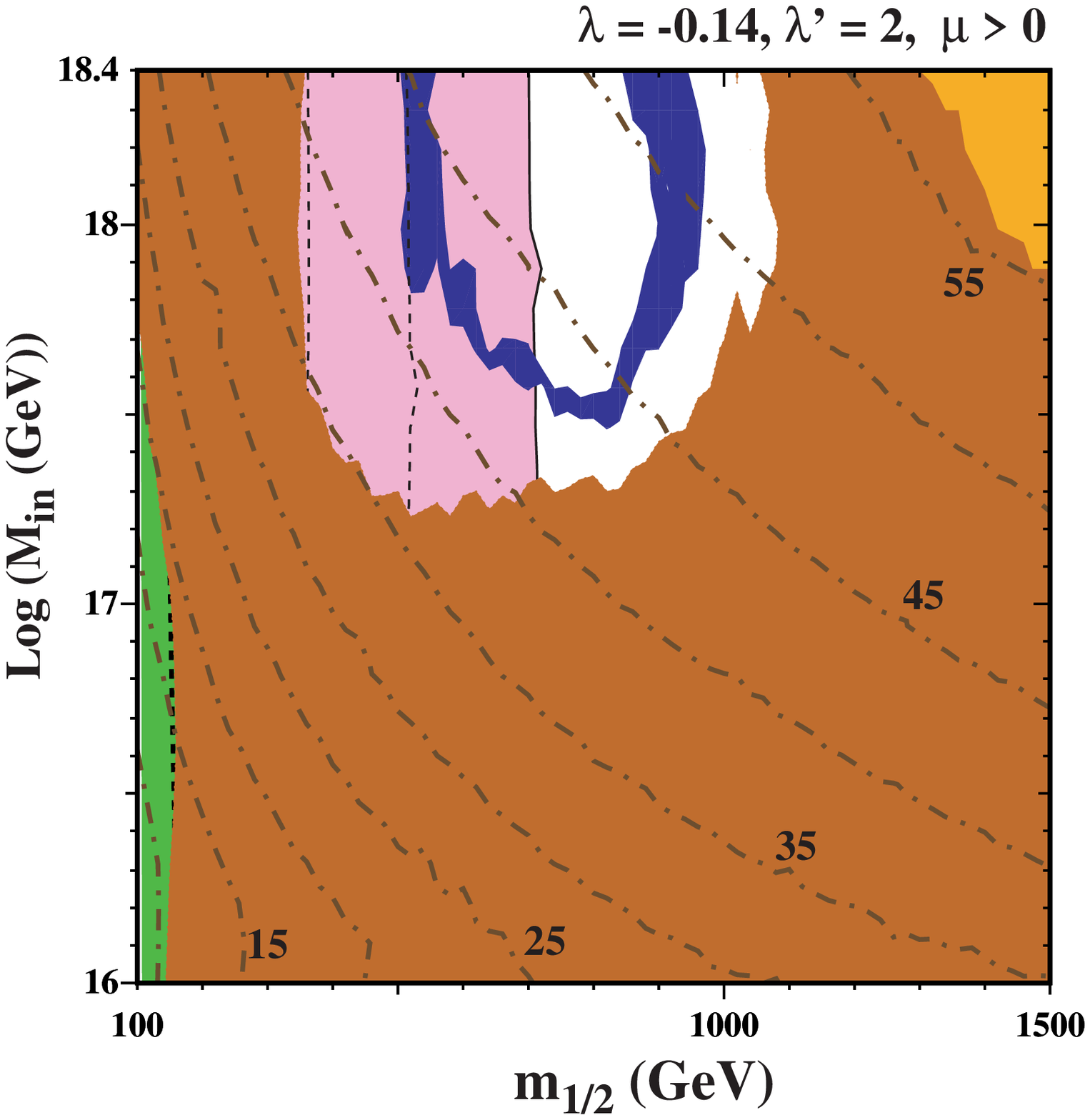} 
    \end{figure}

Panel (b) of Fig.~\ref{noscsgut} for $\lambda = -0.14$ has a rather 
different appearance, but is in fact
a natural continuation of the trends seen in the previous panel. In particular, the 
near-vertical part of the WMAP-compatible strip has moved to
larger $m_{1/2} \sim 400$~GeV, and is compatible with both $m_h$ and
$b \to s \gamma$, and the near-horizontal part of the strip has risen to
$M_{in} \sim 3 \times 10^{17}$~GeV.
More dramatically, the WMAP-compatible strip now becomes a
loop, with a right side connecting the previous two strips at relatively
large $m_{1/2}$ and $M_{in}$ and $\tan \beta \ga 50$
(though the loop closes only when $M_{in} > M_P$).
We emphasize that all of the 
loop has a neutralino LSP, and that the stau-LSP contour surrounds
this loop: the region within the loop has too much dark matter.

In panel (c) of Fig. \ref{noscsgut}
for $\lambda = -0.15$ we see just a `blob' with $m_{1/2} \in (500, 650)$~GeV
and $M_{in} \in (5 \times 10^{17}, 3 \times 10^{18})$~GeV on the edge of the region 
preferred by $g_\mu - 2$. This remaining `blob' 
disappears for larger $-\lambda$ as shown in panel (d) for  $\lambda = -0.16$.
Here, the area within the `window' is phenomenologically allowed, though the
relic density lies below the WMAP range.

In summary, 
we have seen that the CMSSM and mSUGRA are in fact different theories
(despite their interchangeable use in the literature).
In mSUGRA, $\tan \beta$ is generally small and funnel regions
with the correct relic density do not appear.  Focus point regions are also absent
due to either a large value of $A_0/m_0$ or small $\tan \beta$.   
In addition, the gravitino is often the LSP 
in mSUGRA models.  For the specific case, of no-scale supergravity with GUT scale universality,
we have seen that the model is phenomenologically challenged.  For $M_{in} > M_{GUT}$,
no-scale supergravity models can however be resurrected.  An upcoming challenge
will be to differentiate between these models once data from the LHC or direct detection experiments
become available.

%%%%%%%%%%%%%%%%%%%%%%%%%%%%%%%%%%%%%%%%%%%%%%%%
%% BACKMATTER
%%%%%%%%%%%%%%%%%%%%%%%%%%%%%%%%%%%%%%%%%%%%%%%%

\begin{theacknowledgments}
 This work was supported in part by DOE grant DE-FG02-94ER-40823.
\end{theacknowledgments}



\bibliographystyle{aipproc}   % if natbib is available

%\bibliography{sample}

 % \typeout{******************************************}
 % \typeout{** Please run "bibtex \jobname" to optain}
 % \typeout{** the bibliography and then re-run LaTeX}
 % \typeout{** twice to fix the references!}
 % \typeout{******************************************}
 % \typeout{}
% }

\end{document}